\documentclass[final,5p,times,twocolumn]{elsarticle}

\usepackage{graphicx}

\usepackage{amssymb}
\usepackage{amsthm}

\journal{Physics Letters B}

\usepackage{amsmath}
\usepackage{bm}
\usepackage{array}
\usepackage{hyperref}
\usepackage{slashed}
\usepackage{braket}

\begin{document}

\begin{frontmatter}



\title{Quark angular momentum in a spectator model}


\author[PKU]{Tianbo Liu}
\ead{liutb@pku.edu.cn}

\author[PKU,CICQM,CHEP]{Bo-Qiang Ma}
\ead{mabq@pku.edu.cn}

\address[PKU]{School of Physics and State Key Laboratory of Nuclear Physics and Technology, Peking University, Beijing 100871, China}
\address[CICQM]{Collaborative Innovation Center of Quantum Matter, Beijing, China}
\address[CHEP]{Center for High Energy Physics, Peking University, Beijing 100871, China}

\begin{abstract}
We investigate the quark angular momentum in a model with the nucleon being a quark and a spectator. Both scalar and axial-vector spectators are included. We perform the calculations in the light-cone formalism where the parton concept is well defined. We calculate the quark helicity and canonical orbital angular momentum. Then we calculate the gravitational form factors which are often related to the kinetic angular momentums, and find that even in a no gauge field model we cannot identify the canonical angular momentums with half the sum of gravitational form factors. In addition, we examine the model relation between the orbital angular momentum and pretzelosity, and find it is violated in the axial-vector case.
\end{abstract}

\begin{keyword}
proton spin \sep orbital angular momentum \sep gravitational form factor \sep pretzelosity

\end{keyword}

\end{frontmatter}


\section{Introduction}
\label{}

Hadrons are bound states of the strong interaction which is described by the quantum chromodynamics (QCD) in the framework of Yang-Mills gauge field theory. One of the central problems in particle physics is to determine nucleon structures in terms of quark and gluon degrees of freedom. The decomposition of the proton spin is one of the most active frontiers in recent years. Although the total angular momentum of an isolated system is well defined, the decomposition to each constituent of a relativistic composite particle, such as the proton, is non-trivial and of great interest.

The observation that only a small fraction~\cite{Ashman:1987hv,Ashman:1989ig} (about 30\% in recent analysis~\cite{Ageev:2005gh,Alexakhin:2006oza,Airapetian:2006vy}) of the proton spin is carried by quark spins has puzzled the physics community for more than two decades. This result severely deviates from the naive quark model where the proton spin is from quark spins. Many possible ways to understand the ``proton spin crisis" have been proposed, such as   to attribute the remaining proton spin to the orbital angular momentum (OAM) and/or the gluon helicity. Due to the Wigner rotation effect~\cite{Wigner:1939cj} which relates the spinors in different frames, the constituent's spin of a composite particle in the rest frame can be decomposed into a spin part and a non-vanishing OAM in the infinite momentum frame (IMF) or light-cone formalism where the parton language is defined~\cite{Ma:1991xq,Ma:1992sj,Ma:1998ar}. Therefore, the OAM plays an important role in understanding the ``proton spin puzzle'', although the gluon helicity also contributes a large fraction~\cite{deFlorian:2014yva}. However, the decomposition of proton spin, especially the definition of OAM, is still under controversy.

A most intuitive decomposition is to divide the proton spin into quark spin, quark orbit, gluon spin and gluon orbit terms~\cite{Jaffe:1989jz}:
\begin{equation}\label{JM}
S_q+L_q+S_g+L_g=\frac{1}{2},
\end{equation}
where the quark orbit operator is defined as
\begin{equation}
L_q=-i\bar{\psi}\gamma^+\bm{r}\times\nabla\psi.
\end{equation}
But the $L_q$, as well as $S_g$ and $L_g$, is not obviously gauge-invariant and thus renders the physical meanings in common situations obscure. To solve this problem, an explicitly gauge-invariant decomposition is proposed~\cite{Ji:1996ek}:
\begin{equation}\label{Ji}
S_q+L_q'+J_g'=\frac{1}{2},
\end{equation}
where each term is obviously gauge-invariant. It shares the same definition for the quark spin operator in (\ref{JM}), but takes a different definition for the quark orbit operator as
\begin{equation}
L_q'=i\bar{\psi}\gamma^+\bm{r}\times\bm{D}\psi,
\end{equation}
where $\bm{D}=-\nabla-ig\bm{A}$ is the covariant derivative. In this decomposition, the total angular momentum for each parton flavor is usually supposed to be related to the sum of two gravitational form factors:
\begin{equation}\label{jrel}
J_{q/g}'=\frac{1}{2}[A_{q/g}(0)+B_{q/g}(0)],
\end{equation}
where the two form factors $A$ and $B$ can be measured through the deeply virtual Compton scattering (DVCS) process.

Recently, Chen {\it et al.} revived the idea to decompose the gauge potential $A_\mu$ into a pure gauge term $A_\mu^{\textrm{pure}}$, which plays the role on gauge symmetry and only has to do with unphysical degrees of freedom, and a physical term $A_\mu^{\textrm{phy}}$, which involves the two physical degrees of freedom~\cite{Chen:2008ag,Chen:2009mr}. With this approach, many more decomposition versions were proposed~\cite{Wakamatsu:2010cb,Hatta:2011ku,Leader:2011za}. As observed in~\cite{Stoilov:2010pv} and discussed in details in~\cite{Lorce:2012ce,Lorce:2012rr}, this kind of split introduces a so-called Stuekelberg symmetry which copies the group of gauge symmetry but acts on the fields on a different manner. Thus, the approach of Chen {\it et al.} can be viewed as a gauge invariant extension (GIE) based on a Stuekelberg symmetry fixing procedure. This procedure is essentially a choice of the physical term which is frame dependent, and therefore may result in different decomposition versions which actually correspond to different physical objects~\cite{Wakamatsu:2014toa}. Nowadays, all the decompositions are usually classified into two groups~\cite{Leader:2013jra,Wakamatsu:2014zza}, the canonical version and the kinetic (or mechanical) version. Due to the GIE procedure, they are both in principle measurable without gauge-invariance breaking.

In this letter, instead of focusing on the controversy on which version is more physical, we perform a calculation of quark angular momentums in a spectator model with the nucleon to be a struck quark and a spectator. We find that even in this no gluon model we cannot identify the canonical angular momentum with half the sum of two gravitational form factors. Then, we also examine the model relation between the pretzelosity and OAM, and find that the relation is violated in the axial-vector case.

\section{Light-cone spectator model}

Hadrons are the eigenstates of the light-cone Hamiltonian $H_{\textrm{LC}}=2P^+P^--\bm{P}_\perp^2$ with invariant mass square as the eigenvalues. Quantized at a fixed light-cone time $\tau=(t+z)/\sqrt{2}$, one may have unambiguous definition on the constituents, and hence a hadron state can be expanded on a complete basis of Fock states as~\cite{Brodsky:1997de}
\begin{equation}
\begin{split}
|H;P^+,\bm{P}_\perp,S_z\rangle=&\sum_{n,\{\lambda_i\}}\prod_{i=1}^N\int\frac{dx_id^2\bm{k}_{\perp i}}{2\sqrt{x_i}(2\pi)^3}16\pi^3\delta(1-\sum_{j=1}^N x_j)\\
&\times\delta^{(2)}(\sum_{j=1}^N\bm{k}_{\perp j})\psi_{n/H}|n;x_j,\bm{k}_{\perp j},\lambda_j\rangle,
\end{split}
\end{equation}
where $N$ is the number of constituents of the Fock state $|n\rangle$, $x_i$, $\bm{k}_{\perp i}$ and $\lambda_i$ are the light-cone momentum fraction, intrinsic transverse momentum and light-cone helicity carried by the $i$-th constituent respectively. The $\psi_{n/H}$ is the light-cone wave function (LCWF) which decribes the probability amplitude to find the Fock state $|n\rangle$ in the hadron state $|H\rangle$.

In the spectator model, the proton is viewed as a struck quark and a spectator which contains the remaining part. Then the proton state is expressed as
\begin{equation}\label{qD}
|p\rangle=\sum_{q,D,\lambda}\int\frac{dxd^2\bm{k}_\perp}{2(2\pi)^3\sqrt{x(1-x)}}\psi_{qD}(x,\bm{k}_\perp)|qD;x,\bm{k}_\perp,\lambda\rangle,
\end{equation}
where $x$ and $\bm{k}_\perp$ are the light-cone momentum fraction and intrinsic transverse momentum carried by the quark. The $D$ represents the spectator. Constrained by the quantum numbers of the quark and proton, the spectator can only be either a scalar or an axial-vector, and the axial-vector one is necessary for flavor separation. One can effectively introduce the quark-spectator-proton vertex in the Lagrangian as~\cite{Bacchetta:2008af}
\begin{equation}\label{lag}
\mathcal{L}_I=g_{_\textrm{S}}\bar{\Psi}\phi\psi+g_{_\textrm{V}}\bar{\Psi}\gamma^\mu\gamma_5A_\mu\psi+h.c.,
\end{equation}
where $\psi$, $\Psi$, $\phi$ and $A_\mu$ are the operators of quark, proton, scalar diquark and axial-vector diquark fields. Some suitable form factors are included in the effective couplings $g_{_\textrm{S/V}}$ to describe the structures. Then with the Dirac structure, we write down the quark-spectator LCWFs as
\begin{align}
{\psi^{\textrm{S}}}^{\Lambda}_{\lambda}(x,\bm{k}_\perp)&=\frac{\bar{u}(k,\lambda)}{\sqrt{2k^+}}\bm{1}\frac{u(P,\Lambda)}{\sqrt{2P^+}}\phi^{\textrm{S}}(x,\bm{k}_\perp),\label{qS}\\
{\psi^{\textrm{V}}}^{\Lambda}_{\lambda\lambda'}(x,\bm{k}_\perp)&=\frac{\bar{u}(k,\lambda)}{\sqrt{2k^+}}\epsilon^*_\mu(p,\lambda')\gamma^\mu\gamma_5\frac{u(P,\Lambda)}{\sqrt{2P^+}}\phi^{\textrm{V}}(x,\bm{k}_\perp),\label{qV}
\end{align}
where $\lambda$, $\lambda'$ and $\Lambda$ are the light-cone helicities of the quark, axial-vector spectator and proton, and the superscripts $\textrm{S}$ and $\textrm{V}$ denote the type of the spectator. The $k$, $p$ and $P$ are the momentums carried by the quark, spectator and proton respectively. For the Dirac spinor $u$ and polarization vector $\epsilon$, we adopt the Lepage-Brodsky convention~\cite{Lepage:1980fj}. The $\phi(x,\bm{k}_\perp)$ is a spin-averaged momentum space wave function. In the following calculations, we choose the form as~\cite{Hwang:2007tb,Liu:2014vwa}
\begin{equation}\label{HM}
\phi^{\textrm{S/V}}(x,\bm{k}_\perp)=\frac{g_{_\textrm{S/V}}}{x\sqrt{1-x}}\left(M^2-\frac{m^2+\bm{k}_\perp^2}{x}-\frac{M_D^2+\bm{k}_\perp^2}{1-x}\right)^{-2},
\end{equation}
where $m$, $M_D$ and $M$ are the masses of the quark, spectator and proton, and $g_{_\textrm{S/V}}$ is a coupling constant. This choice corresponds to an effective coupling in the Lagrangian as $g_{_\textrm{S/V}}(k^2)=g_{_\textrm{S/V}}/(k^2-m^2)$. It respects the Lorentz invariance~\cite{Brodsky:2003pw} and leads to the polynomiality property of generalized parton distribution (GPD) moments~\cite{Hwang:2007tb}. Replacing the quark mass $m$ by a cut-off parameter, one can get the form induced by a dipole form factor~\cite{Jakob:1997wg,Bacchetta:2008af}. The LCWFs in (\ref{qS}) and (\ref{qV}) are normalized as
\begin{equation}
\begin{split}
&\int\frac{dxd^2\bm{k}_\perp}{16\pi^3}\sum_{\lambda,\lambda'}|{\psi^{\textrm{S/V}}}^{\Lambda}_{\lambda\lambda'}(x,\bm{k}_\perp)|^2\\
=&\int\frac{dxd^2\bm{k}_\perp}{16\pi^3}N_{\textrm{S/V}}^2|\phi^{\textrm{S/V}}(x,\bm{k}_\perp)|^2=1,
\end{split}
\end{equation}
where $N_{\textrm{S/V}}$ is the normalization factor for spin states:
\begin{align}
N_{\textrm{S}}^2=&\frac{(m+xM)^2+\bm{k}_\perp^2}{x^2},\\
\begin{split}\label{vnor}
N_{\textrm{V}}^2=&\frac{2(1+x^2)\bm{k}_\perp^2+2(1-x)^2(m+xM)^2}{x^2(1-x)^2}\\
&+\frac{(\bm{k}_\perp^2-xM_D^2-(1-x)^2mM)^2}{x^2(1-x)^2M_D^2}\\
&+\frac{(m+M)^2\bm{k}_\perp^2}{x^2M_D^2}.
\end{split}
\end{align}

\section{Quark angular momentum and gravitational form factors}

In the quark-spectator state expansion (\ref{qD}), the Fock states $|qD\rangle$ are the eigenstates of quark spin $S_q$ and total OAM $L$. Thus, we can express the expected value of quark spin in a polarized proton as
\begin{equation}
\begin{split}\label{ss}
S_q^{\textrm{S}}&=\frac{1}{2}\int\frac{dxd^2\bm{k}_\perp}{16\pi^3}\left[|{\psi^{\textrm{S}}}^{\uparrow}_\uparrow(x,\bm{k}_\perp)|^2-|{\psi^{\textrm{S}}}^{\uparrow}_\downarrow(x,\bm{k}_\perp)|^2\right]\\
&=\frac{1}{2}\int\frac{dxd^2\bm{k}_\perp}{16\pi^3}W^{\textrm{S}}_S(x,\bm{k}_\perp)N_{\textrm{S}}^2|\phi^{\textrm{S}}(x,\bm{k}_\perp)|^2,
\end{split}
\end{equation}
where
\begin{equation}
W^{\textrm{S}}_S(x,\bm{k}_\perp)=\frac{(m+xM)^2-\bm{k}_\perp^2}{(m+xM)^2+\bm{k}_\perp^2},
\end{equation}
in the scalar case, and
\begin{equation}
\begin{split}\label{sv}
S_q^{\textrm{V}}&=\frac{1}{2}\int\frac{dxd^2\bm{k}_\perp}{16\pi^3}\sum_{\lambda'}\left[|{\psi^{\textrm{V}}}^{\uparrow}_{\uparrow\lambda'}(x,\bm{k}_\perp)|^2-|{\psi^{\textrm{V}}}^{\uparrow}_{\downarrow\lambda'}(x,\bm{k}_\perp)|^2\right]\\
&=\frac{1}{2}\int\frac{dxd^2\bm{k}_\perp}{16\pi^3}W^{\textrm{V}}_S(x,\bm{k}_\perp)N_{\textrm{V}}^2|\phi^{\textrm{V}}(x,\bm{k}_\perp)|^2,
\end{split}
\end{equation}
where
\begin{equation}
\begin{split}
W^{\textrm{V}}_S(x,\bm{k}_\perp)=&\frac{2(1+x^2)\bm{k}_\perp^2-2(1-x)^2(m+xM)^2}{N_{\textrm{V}}^2x^2(1-x)^2}\\
&+\frac{(\bm{k}_\perp^2-xM_D^2-(1-x)^2mM)^2}{N_{\textrm{V}}^2x^2(1-x)^2M_D^2}\\
&-\frac{(m+M)^2\bm{k}_\perp^2}{N_{\textrm{V}}^2 x^2M_D^2},
\end{split}
\end{equation}
in the axial-vector case. The $W$ factors reflect relativistic effects, such as the Wigner rotation effect. With the LCWFs in (\ref{HM}), (\ref{ss}) and (\ref{sv}) are analytically integrable. Substituting the quark and spectator mass parameters which are fitted to the electromagnetic form factors in~\cite{Liu:2014vwa}, we get the numerical values as
\begin{align}
S_q^{\textrm{S}}&=0.383,\\
S_q^{\textrm{V}}&=-0.135.
\end{align}

Similar to the quark spin, the total OAM can be expressed as
\begin{align}
\begin{split}
L^{\textrm{S}}&=\int\frac{dxd^2\bm{k}_\perp}{16\pi^3}|{\psi^{\textrm{S}}}^{\uparrow}_\downarrow(x,\bm{k}_\perp)|^2\label{ls}\\
&=\int\frac{dxd^2\bm{k}_\perp}{16\pi^3}W^{\textrm{S}}_L(x,\bm{k}_\perp)N_{\textrm{S}}^2|\phi^{\textrm{S}}(x,\bm{k}_\perp)|^2,
\end{split}\\
\begin{split}\label{lv}
L^{\textrm{V}}&=\int\frac{dxd^2\bm{k}_\perp}{16\pi^3}\big[-|{\psi^{\textrm{V}}}^{\uparrow}_{\uparrow+}(x,\bm{k}_\perp)|^2+|{\psi^{\textrm{V}}}^{\uparrow}_{\uparrow-}(x,\bm{k}_\perp)|^2\\
&\quad+|{\psi^{\textrm{V}}}^{\uparrow}_{\downarrow0}(x,\bm{k}_\perp)|^2+2|{\psi^{\textrm{V}}}^{\uparrow}_{\downarrow-}(x,\bm{k}_\perp)|^2\big]\\
&=\int\frac{dxd^2\bm{k}_\perp}{16\pi^3}W^{\textrm{V}}_L(x,\bm{k}_\perp)N_{\textrm{V}}^2|\phi^{\textrm{V}}(x,\bm{k}_\perp)|^2,
\end{split}
\end{align}
where the $W$ factors are
\begin{align}
W^{\textrm{S}}_L(x,\bm{k}_\perp)=&\frac{\bm{k}_\perp^2}{(m+xM)^2+\bm{k}_\perp^2},\\
\begin{split}
W^{\textrm{V}}_L(x,\bm{k}_\perp)=&\frac{(m+M)^2\bm{k}_\perp^2}{N_{\textrm{V}}^2x^2M_D^2}-\frac{2(1-x^2)\bm{k}_\perp^2}{N_{\textrm{V}}^2x^2(1-x)^2}.
\end{split}
\end{align}
The $L^{\textrm{S}}$ and $L^{\textrm{V}}$ are expected values of the total intrinsic OAM operator
\begin{equation}\label{ioam}
\hat{L}=-i\sum_{i=1}^{N-1}\bm{k}_{\perp i}\times\frac{\partial}{\partial\bm{k}_{\perp i}}.
\end{equation}
Among $N$ intrinsic transverse momentums, only $N-1$ of them are independent. To get the canonical OAM carried by the quark, one may evaluate the expected value of the intrinsic OAM operator with respect to the transverse center~\cite{Burkardt:2000za,Lorce:2011kn}. For the quark-spectator system, the quark intrinsic OAM can be evaluated from~\cite{Harindranath:1998ve}
\begin{equation}
-i(1-x)\bm{k}_\perp\times\frac{\partial}{\partial\bm{k}_\perp}.
\end{equation}
The numerical values are
\begin{align}
L_q^{\textrm{S}}&=0.089,\\
L_q^{\textrm{V}}&=0.017.
\end{align}
Then the quark total canonical angular momentums are
\begin{align}
J_q^{\textrm{S}}&=S_q^{\textrm{S}}+L_q^{\textrm{S}}=0.472,\\
J_q^{\textrm{V}}&=S_q^{\textrm{V}}+L_q^{\textrm{V}}=-0.118.
\end{align}
Including the angular momentum carried by the spectator, the angular momentum sum rule is satisfied in both the scalar and the axial-vector cases:
\begin{align}
J^{\textrm{S}}&=S_q^{\textrm{S}}+L_q^{\textrm{S}}+L_D^{\textrm{S}}=\frac{1}{2},\\
J^{\textrm{V}}&=S_q^{\textrm{V}}+L_q^{\textrm{V}}+S_D^{\textrm{V}}+L_D^{\textrm{V}}=\frac{1}{2}.
\end{align}
In this no gluon model, these values are expected to be equal to the kinetic angular momentums, since the difference between the two definitions is an interaction term with the gluon.

However, with the relation (\ref{jrel}), it is suggested to obtain the kinetic angular momentums for each constituent from the sum of two gravitational form factors $A(Q^2)$ and $B(Q^2)$, which can be measured through the DVCS process. Similar to the Dirac and Pauli form factors, these two gravitational form factors can be calculated from the helicity-conserved and helicity-flip matrix elements of the energy momentum tensor current as~\cite{Brodsky:2000ii,Liu:2014npa}
\begin{align}
\langle P+q,\uparrow|T^{++}(0)|P,\uparrow\rangle&=2(P^+)^2A(Q^2),\\
\langle P+q,\uparrow|T^{++}(0)|P,\downarrow\rangle&=2(P^+)^2\frac{-(q^1-iq^2)}{2M}B(Q^2),
\end{align}
where $q^2=-Q^2$ is transfered momentum square. Using the Noether theorem~\cite{Noether:1918zz} and the effective Lagrangian in (\ref{lag}), we can derive the canonical energy momentum tensor, which differs from the Belinfante improved energy momentum tensor by an antisymmetric term, as
\begin{align}
T^{\mu\nu}_q=&\frac{i}{2}[\bar{\psi}\gamma^\mu\partial^\nu\psi-(\partial^\nu\bar{\psi})\gamma^\mu\psi]-g^{\mu\nu}\mathcal{L}_q,\\
T^{\mu\nu}_D=&\partial^\mu\phi\partial^\nu\phi-F^{\mu\rho}\partial^\nu A_\rho-g^{\mu\nu}(\mathcal{L}_{\textrm{S}}+\mathcal{L}_{\textrm{V}}),\\
T^{\mu\nu}_I=&-g^{\mu\nu}\mathcal{L}_I,
\end{align}
where the subscripts $q$, $D$ and $I$ denote the quark, spectator and interaction parts respectively, and $\mathcal{L}_{q/\textrm{S/V}}$ represents the Lagrangian of free quark, scalar and vector fields. Taking the plus-plus component of the energy momentum tensor, one can easily find that the contribution from the interaction term vanishes. Thus the contributions from the quark and the spectator are separated as
\begin{align}
T^{++}_q=&\frac{i}{2}[\bar{\psi}\gamma^+\partial^+\psi-(\partial^+\bar{\psi})\gamma^+\psi],\\
T^{++}_D=&\partial^+\phi\partial^+\phi-F^{+\rho}\partial^+A_\rho,
\end{align}
where the expression for the quark park is the same as that in QCD with the light-cone gauge $A_g^+=0$.

After some algebra, the quark parts of these two form factors are expressed in terms of the overlap of LCWFs as
\begin{align}
A^{\textrm{S}}_q(Q^2)&=\int\frac{dxd^2\bm{k}_\perp}{16\pi^3}\sum_\lambda x{\psi^{\textrm{S}}}^{\uparrow*}_\lambda(x,\bm{k}_\perp'){\psi^{\textrm{S}}}^{\uparrow}_\lambda(x,\bm{k}_\perp),\\
B^{\textrm{S}}_q(Q^2)&=-\frac{2M}{q^1-iq^2}\int\frac{dxd^2\bm{k}_\perp}{16\pi^3}\sum_\lambda x{\psi^{\textrm{S}}}^{\uparrow*}_\lambda(x,\bm{k}_\perp'){\psi^{\textrm{S}}}^{\downarrow}_\lambda(x,\bm{k}_\perp),
\end{align}
for the scalar case, and
\begin{align}
A^{\textrm{V}}_q(Q^2)&=\int\frac{dxd^2\bm{k}_\perp}{16\pi^3}\sum_{\lambda,\lambda'} x{\psi^{\textrm{V}}}^{\uparrow*}_{\lambda\lambda'}(x,\bm{k}_\perp'){\psi^{\textrm{V}}}^{\uparrow}_{\lambda\lambda'}(x,\bm{k}_\perp),\\
B^{\textrm{V}}_q(Q^2)&=-\frac{2M}{q^1-iq^2}\int\frac{dxd^2\bm{k}_\perp}{16\pi^3}\sum_{\lambda,\lambda'} x{\psi^{\textrm{V}}}^{\uparrow*}_{\lambda\lambda'}(x,\bm{k}_\perp'){\psi^{\textrm{V}}}^{\downarrow}_{\lambda\lambda'}(x,\bm{k}_\perp),
\end{align}
for the axial-vector case. The $\bm{k}_\perp'=\bm{k}_\perp+(1-x)\bm{q}_\perp$ is quark intrinsic transverse momentum in the final state. At $Q^2=0$, they can be expressed as
\begin{align}
A^{\textrm{S}}_q(0)=&\int\frac{dxd^2\bm{k}_\perp}{16\pi^3}W^{\textrm{S}}_A(x,\bm{k}_\perp)N_{\textrm{S}}^2x|\phi^{\textrm{S}}(x,\bm{k}_\perp)|^2,\\
B^{\textrm{S}}_q(0)=&\int\frac{dxd^2\bm{k}_\perp}{16\pi^3}W^{\textrm{S}}_B(x,\bm{k}_\perp)N_{\textrm{S}}^2x|\phi^{\textrm{S}}(x,\bm{k}_\perp)|^2,
\end{align}
where
\begin{align}
W^{\textrm{S}}_A(x,\bm{k}_\perp)=&1,\\
W^{\textrm{S}}_B(x,\bm{k}_\perp)=&\frac{2M(1-x)(m+xM)}{(m+xM)^2+\bm{k}_\perp^2},
\end{align}
and
\begin{align}
A^{\textrm{V}}_q(0)=&\int\frac{dxd^2\bm{k}_\perp}{16\pi^3}W^{\textrm{V}}_A(x,\bm{k}_\perp)N_{\textrm{V}}^2x|\phi^{\textrm{V}}(x,\bm{k}_\perp)|^2,\\
B^{\textrm{V}}_q(0)=&\int\frac{dxd^2\bm{k}_\perp}{16\pi^3}W^{\textrm{V}}_B(x,\bm{k}_\perp)N_{\textrm{V}}^2x|\phi^{\textrm{V}}(x,\bm{k}_\perp)|^2,
\end{align}
where
\begin{align}
W^{\textrm{V}}_A=&1\\
\begin{split}
W^{\textrm{V}}_B=&\frac{2M(m+M)(\bm{k}_\perp^2-xM_D^2-(1-x)^2mM)}{N_{\textrm{V}}^2x^2M_D^2}\\
&-\frac{4M(m+xM)}{N_{\textrm{V}}^2x}.
\end{split}
\end{align}
The numerical values are
\begin{align}
A_q^{\textrm{S}}(0)&=0.290,\\
B_q^{\textrm{S}}(0)&=0.422,\\
A_q^{\textrm{V}}(0)&=0.294,\\
B_q^{\textrm{V}}(0)&=-0.370.
\end{align}
Including the contributions from the spectator, the momentum sum rule:
\begin{equation}
A^{\textrm{S/V}}(0)=A_q^{\textrm{S/V}}(0)+A_D^{\textrm{S/V}}(0)=1,
\end{equation}
and the anomalous gravitomagnetic moment sum rule~\cite{Teryaev:1999su}:
\begin{equation}
B^{\textrm{S/V}}(0)=B_q^{\textrm{S/V}}(0)+B_D^{\textrm{S/V}}(0)=0,
\end{equation}
are satisfied in both cases.

Therefore, the total angular momentum is equal to half the sum of these two form factors
\begin{equation}
J^{\textrm{S/V}}=\frac{1}{2}[A^{\textrm{S/V}}(0)+B^{\textrm{S/V}}(0)].
\end{equation}
But, comparing the quark part with the quark canonical angular momentums calculated above, we find that even in such a no gluon model, where the difference between the canonical and kinetic operators $L_q$ and $L_q'$ has no contributions, one cannot identify the canonical angular momentums with half the sum of two gravitational form factors for each constituent:
\begin{align}
\frac{1}{2}[A_q^{\textrm{S}}(0)+B_q^{\textrm{S}}(0)]&=0.356\neq J_q^{\textrm{S}},\\
\frac{1}{2}[A_q^{\textrm{V}}(0)+B_q^{\textrm{V}}(0)]&=-0.038\neq J_q^{\textrm{V}}.
\end{align}
In other words, the relation (\ref{jrel}) is violated in this model, in both the scalar and the axial-vector cases.

Here we specified a form, the Hwang-Mueller prescription~\cite{Hwang:2007tb,Liu:2014vwa}, for the spin-averaged LCWFs $\phi^{\textrm{S/V}}(x,\bm{k}_\perp)$ to get the numerical values quantitatively. This form respects the Lorentz invariance and produces the GPD polynomiality property. Technically, it makes all the integrals in this letter analytically integrable. Apart from this prescription, there are many other choices, such as the Brodsky-Huang-Lepage (BHL) prescription~\cite{Brodsky:1980vj,Brodsky:1981jv,Brodsky:1982nx}, the Terentev-Karmanov (TK) prescription~\cite{Terentev:1976jk,Karmanov:1979if}, the Chung-Coester-Polyzou (CCP) prescription~\cite{Chung:1988mu}, the Vega-Schmidt-Gutsche-Lyubovitskij (VSGL) prescription~\cite{Vega:2013bxa} and so on. These prescriptions have nothing to do with any $W$ factors which are determined by Dirac structures. Therefore, any choice will not change any conclusion in this letter, although it will indeed change the numerical values quantitatively.

\section{Pretzelosity and orbital angular momentum}

The pretzelosity, denoted as $h_{1T}^\perp$, is one of the eight leading twist transverse momentum dependent parton distributions (TMDs). It represents the probability to find a transverse polarized quark in a perpendicularly transverse polarized proton, and can be measured through the single spin asymmetries in the semi-inclusive deep inelastic scattering (SIDIS) process~\cite{Kotzinian:1994dv,Bacchetta:2006tn}. Based on some model calculations, it is suggested to relate the pretzelosity to the OAM~\cite{She:2009jq,Lorce:2011kn}.

In the scalar case, the pretzelosity is expressed as
\begin{equation}
\begin{split}
h_{1T}^{\perp\textrm{S}}(x,\bm{k}_\perp)&=-\frac{1}{16\pi^3}\frac{2M^2}{x^2}|\phi^{\textrm{S}}(x,\bm{k}_\perp)|^2\\
&=-\frac{N_{_\textrm{S}}^2}{16\pi^3}\frac{2M^2(1-x)^3}{[\bm{k}_\perp^2+\Lambda^2]^4},
\end{split}
\end{equation}
where
\begin{equation}
\Lambda^2=xM_D^2+(1-x)m^2-x(1-x)M^2.
\end{equation}
Its first transverse momentum moment is equal to the OAM of the quark-spectator system with a minus sign~\cite{She:2009jq}:
\begin{equation}\label{pret}
L^{\textrm{S}}=-\int dxd^2\bm{k}_\perp\frac{\bm{k}_\perp^2}{2M^2}h_{1T}^{\perp\textrm{S}}(x,\bm{k}_\perp)=0.117,
\end{equation}
which is also equal to the difference between the helicity and transversity~\cite{Ma:1998ar,Bacchetta:2008af}. Similarly, the intrinsic OAM carried by the quark can be expressed with the pretzelosity as
\begin{equation}\label{qpret}
L_q^{\textrm{S}}=-\int dxd^2\bm{k}_\perp\frac{\bm{k}_\perp^2}{2M^2}(1-x)h_{1T}^{\perp\textrm{S}}(x,\bm{k}_\perp).
\end{equation}

However, in the axial-vector case, the pretzelosity is expressed as
\begin{equation}
\begin{split}\label{vv}
h_{1T}^{\perp\textrm{V}}(x,\bm{k}_\perp)&=-\frac{1}{16\pi^3}\frac{2M^2(m+M)^2}{x^2M_D^2}|\phi^{\textrm{V}}(x,\bm{k}_\perp)|^2\\
&=-\frac{N_{_\textrm{V}}^2}{16\pi^3}\frac{2M^2(m+M)^2(1-x)^3}{M_D^2[\bm{k}_\perp^2+\Lambda^2]^4}.
\end{split}
\end{equation}
Its first transverse momentum moment is
\begin{equation}\label{vpret}
\int dxd^2\bm{k}_\perp\frac{\bm{k}_\perp^2}{2M^2}h_{1T}^{\perp\textrm{V}}=-\int\frac{dxd^2\bm{k}_\perp}{16\pi^3}|{\psi^{\textrm{V}}}^{\uparrow}_{\downarrow0}(x,\bm{k}_\perp)|^2,
\end{equation}
which is equal to the third term of the $L^{\textrm{V}}$ in (\ref{lv}) with a minus sign. Hence, simple extensions of the relations (\ref{pret}) and (\ref{qpret}) to the axial-vector case are not justified, because in this case the model is not spherically symmetric as demonstrated in~\cite{Lorce:2011kn,Lorce:2011kd}. If only the transverse polarizations for the axial-vector are included, the pretzelosity will vanish in this model, but the intrinsic OAM is still non-vanishing. Therefore, without further assumptions, there is no general relations between the OAM and pretzelosity. This is consistent with our intuition, since the OAM is essentially a correlation between the coordinate and the momentum, while the pretzelosity, as a leading twist TMD, only contains the momentum information. If the term in (\ref{vpret}) plays the dominant role in $L^{\textrm{V}}$, the relations (\ref{pret}) and (\ref{qpret}) might be extended to the axial-vector case as approximate relations with some correction factors phenomenologically. Comparing (\ref{vv}) with the OAM expression in (\ref{lv}), this correction factor is written as
\begin{equation}
C^{\textrm{V}}(x,\bm{k}_\perp)=\frac{(1-x)(m+M)^2-2(1+x)M_D^2}{(1-x)(m+M)^2}.
\end{equation}

\section{Conclusions}

In this letter, we investigate the quark angular momentum in a spectator model. The calculations are performed in the light-cone formalism where the parton concept is well defined.

Nowadays, there are many decomposition versions for the proton spin. All these decomposition versions are usually classified into two groups, the canonical version and the kinetic version, and the main difference between them is the definition of the OAM~\cite{Leader:2013jra,Wakamatsu:2014zza}. With the GIE procedure, both of them are in principle measurable without gauge-invariance breaking. In this study, regardless of the dispute on which one is more physical, we perform the calculations in a no gluon model where the results from these two definitions are expected to be the same.

Considering the Dirac structure, we write down the LCWFs of the quark-spectator system with the scalar and axial-vector couplings. Then we calculate the spin and intrinsic canonical OAM carried by the quark in both cases. By including the contributions from the spectator, the angular momentum sum rule is satisfied. Taking the relation (\ref{jrel}) as an assumption~\cite{Ji:1996ek}, we calculate the so-called kinetic quark angular momentum through two gravitational form factors. As a direct result of the momentum fraction and the anomalous gravitomagnetic moment sum rules, the total angular momentum is equal to half the sum of these two form factors. However, for each constituent, we cannot identify half the sum of the form factors with the canonical angular momentums, which are expected to be the same as the kinetic ones in a no gluon model, in either the scalar case or the axial-vector case. In other words, the relation (\ref{jrel}) is violated in this model, even though no gluon degrees of freedom are introduced.

In principle, the proton spin decomposition should be understood with explicit calculations in QCD. But due to the nonperturbative nature of QCD at hadron scale, it is almost impossible to perform such an example at present. Instead, an example in QED, which has very similar operator structure to that in QCD up to a color factor, was performed recently, but surprisingly the relation (\ref{jrel}) failed to match order by order in perturbative calculations~\cite{Liu:2015}. An explicit calculation in this letter indicates that neither does the ``general'' relation between the kinetic angular momentum and the gravitational form factors satisfy in the spectator model, which has been widely used to investigate the nucleon structure phenomenologically. Therefore we need more careful scrutiny concerning issues related to the orbital angular momentum of a composite system.

\section*{Acknowledgements}

One of the authors (T. L.) would like to thank C\'edric Lorc\'e for helpful discussions on the definition of the LCWF. This work is supported by the National Natural Science Foundation of China (Grants No. 11035003, No. 11120101004 and No. 11475006).

\appendix

\section{Axial-vector with only transverse polarizations}

If only transverse polarizations, {\it i.e.} $\lambda'=\pm1$, of the axial-vector spectator are included, the spin states normalization factor in (\ref{vnor}) is expressed as
\begin{equation}
N_{\textrm{V}}^2=\frac{2(1+x^2)\bm{k}_\perp^2+2(1-x)^2(m+xM)^2}{x^2(1-x)^2}.
\end{equation}
Such situation applies when the spectator to be massless. Then the $W$ factors are correspondingly written as
\begin{align}
W^{\textrm{V}}_S(x,\bm{k}_\perp)=&\frac{(1+x^2)\bm{k}_\perp^2-(1-x)^2(m+xM)^2}{(1+x^2)\bm{k}_\perp^2+(1-x)^2(m+xM)^2},\\
W^{\textrm{V}}_L(x,\bm{k}_\perp)=&-\frac{(1-x^2)\bm{k}_\perp^2}{(1+x^2)\bm{k}_\perp^2+(1-x)^2(m+xM)^2},\\
W^{\textrm{V}}_A(x,\bm{k}_\perp)=&1,\\
W^{\textrm{V}}_B(x,\bm{k}_\perp)=&-\frac{2M(m+xM)x(1-x)^2}{(1+x^2)\bm{k}_\perp^2+(1-x)^2(m+xM)^2}.
\end{align}






\end{document}